\documentstyle[12pt,aasms4]{article}
\def\lap{\lower.5ex\hbox{$\; \buildrel < \over \sim \;$}}
\def\gap{\lower.5ex\hbox{$\; \buildrel > \over \sim \;$}}
\begin{document}
\title{Resolving the Helium Lyman-$\alpha$ Forest:  Mapping
Intergalactic
Gas and Ionizing Radiation at $z\approx 3$\altaffilmark{1}}
\author{Craig J. Hogan, Scott F. Anderson and Martin H. Rugers}
\affil{University of Washington}
\affil{Astronomy Dept. Box 351580, Seattle, WA  98195-1580}
\affil{hogan@astro.washington.edu, anderson@astro.washington.edu}

\altaffiltext{1}{Based on observations with the NASA/ESA Hubble Space Telescope 
obtained at the Space Telescope Science Institute, which is operated by the 
Association of Universities for Research in Astronomy, Inc., under NASA
contract NAS5-26555.}

\begin{abstract}
We present a new, high resolution HST/GHRS spectrum of quasar Q0302-003,
and use the $He^+$ Lyman-$\alpha$ absorption, together
with a high resolution Keck spectrum of the $HI$
Lyman-$\alpha$ forest,
to probe the distribution and ionization state
of  foreground gas 
just below  the quasar redshift $z\approx 3.3$.
Within $\approx 4000$ km/sec of  the quasar redshift the
spectrum shows a substantial   flux ($\tau \approx 1$) with
``$He^+$ Lyman-$\alpha$ forest''   absorption features
correlated in redshift with the $HI$ Lyman-$\alpha$ forest;
the absorption in this region 
is accounted for entirely
by the discrete components of the forest,
indeed  the main ``Gunn-Peterson edge'' can be identified
with a particular complex of HI absorbing clouds. 
We attribute the lack of continuous absorption from diffuse gas 
to the ``proximity effect'' in this region,
a  large   bubble where helium is highly ionized by  the quasar, and use its
size to estimate the background flux at the $He^+$ 
ionization threshold.
The near-quasar data also lead to constraints on diffuse
gas density  near the quasar, tied to the observed quasar flux,
and helium abundance, tied to the observed quasar spectrum.
Far from the quasar redshift, the spectrum
displays $He^+$  absorption ($\tau\ge 1.3$)
even in several redshift intervals with no detectable HI absorption,
implying   a soft ionizing spectrum
as well as absorption from gas between detected HI clouds.
The smoothed  spectrum displays residual flux everywhere with
an average optical depth $\tau_{GP}\le2$, which
indicates a low   density of redshift-space-filling gas;
using constraints from the HI ionizing spectrum
we estimate  $\Omega_g\le 0.01(h/0.7)^{-1.5}$, and infer
that the helium is already mostly doubly ionized by
this epoch.  Our estimates are consistent
with ionization models based on observed quasar populations,
previous limits from HI Gunn-Peterson studies,
and  simulations of the gas distribution in CDM models
of galaxy formation.   
  
\end{abstract}

\section{Introduction}

The line of sight to the $z=3.285$ quasar Q0302-003 (hereafter, Q0302)
provides a unique probe of the
intergalactic medium  at high redshift. It is the highest redshift
quasar yet discovered where the light is
unobscured down to below 304 \AA\ in the rest frame, allowing measurement of  
$He^+$  Lyman-$\alpha$ absorption by foreground gas. Besides confirming the 
abundant primordial helium predicted by the Big Bang model, the
helium absorption records new information about
the ionization history of intergalactic gas, especially useful
for disentangling the roles of stars and quasars in reionizing the universe.
Most significantly, its relatively
higher optical depth  provides   a better
tool than HI  for measuring absorption by
diffuse gas which fills 
 the space between the galaxies and between
the  identified HI ``Lyman-$\alpha$ forest''
clouds, in the intergalactic and protogalactic media.
Resolved helium absorption provides a direct 
``nonlinear map'' of the gas distribution in
the most rarefied gas occupying most of the
volume of space. 

Ideas about the cosmic gas distribution have sharpened 
quantitatively in recent years, due to hydrodynamic
simulations  which accurately 
predict the motion of matter in  
hierarchical models of galaxy formation
(Cen et al. 1994, Hernquist et al. 1995, Miralda-Escud\'e et al.  1996,
 Rauch et al. 1997,
Croft et al. 1997, Zhang et al. 1997; see also Bi and 
Davidsen 1997).  Departing from earlier 
models based on isolated clouds with symmetric geometries such
as spheres and slabs,   simulations of the conversion 
of uniform gas into condensations reveal a dynamical  system
with a complex geometry where
the distinction between diffuse gas and clouds is blurred
and is not always reflected in the appearance of an absorption spectrum. 
Simulated spectra reveal that
gas  in the most underdense regions,
filling the bulk of the spatial volume,
 is so highly ionized that it
produces  absorption features  with  very low HI Lyman-$\alpha$ optical depth. 
The most  abundant ion ${\rm He^+}$  however produces
optical depths of the order of unity even in these regions, so its absorption 
is easily detectable, mapping  the distribution of
cosmic baryons  at the lowest densities.

Absorption by ${\rm He^+}$ is also
the most direct probe of the hard ultraviolet 
cosmic radiation field, which can be predicted from
semiempirical models based on observed quasar
and absorber  populations
(Haardt and Madau 1996).  The spectral shape also influences
other observables such as the ratio of CIV to SiIV
(Songaila \& Cowie 1996, Giroux \& Shull 1997,
Savaglio et al 1997), so information
from helium absorption allows information about
relative C and Si abundances to be derived.  In situations
where the ionizing spectrum is known, such as the near
proximity of a quasar,  ${\rm He^+}$ absorption can be compared
to HI absorption to extract
independent information about the primordial abundance of helium,
an important test of Big Bang Nucleosynthesis.

The first  detection of cosmic $He^+$ absorption was made  in Q0302 by
Jakobsen et al.
using the Hubble Space Telescope Faint Object Camera (FOC).  They found an
absorption
edge and a large ``Gunn-Peterson'' continuous optical depth, $\tau>1.7$,
attributed to $He^+$ Lyman-$\alpha$  absorption by diffuse gas.  A similar
observation has also been
made of the $z=3.185$ quasar PKS 1935-692, with a similar result
(a lower limit on the optical depth $\tau> 1.5$, also at 90\% confidence,
by Tytler \& Jakobsen 1996; note that this result is based on new   data which
modify earlier conclusions on this object by Tytler et al. 1995).
These data however had important limitations;
the low (10 \AA) resolution of the  FOC  
could not resolve features from the known HI clouds
or even from  the larger gaps between them, 
and the inaccuracy of the calibration (20 \AA) 
could not place the location of the edge accurately   relative
to either  the quasar or the clouds.

A significant improvement came from
  the Hopkins Ultraviolet Telescope (Davidsen et al. 1996),
which can reach shorter wavelengths and hence
lower redshift than HST, and also provides better 
resolution and wavelength calibration than FOC.
Davidsen et al. observed     the $z=2.72$
quasar  HS 1700+64 and found an $He^+$ edge   close enough
to the predicted redshift to rule out the possibility
of foreground HI as an important contaminant.
They also found that the flux below the edge is not
consistent with zero, and measured accurately a
mean optical depth, $\tau=1.00\pm 0.07$.
The decreasing absorption with time  reflects   the increasing ionization
of ${\rm He^+}$  
  at around $z=3$  and   the conversion
of diffuse gas into clouds.

 Our new observations of Q0302 were made
to improve both the wavelength calibration and
resolution of the ${\rm He^+}$ absorption,
with enough sensitivity to correlate usefully 
with the HI absorption. 
This is much more informative than just detecting  the mean
absorption--- we can
explore the relative contributions of clouds
and diffuse gas, as well as measuring  independently the
ionizing radiation field  and the helium abundance.
A substantial optical depth, on the order of the whole
effect detected at low resolution (Songaila et al. 1995),
is expected just from the gas   accounted for in the
discrete clouds already identified as the HI Lyman-$\alpha$ forest.
Although this possibility can be modeled theoretically
in a statistical way for low resolution data (Giroux et al. 1995),
our higher resolution ${\rm He^+}$ spectrum allows a direct
detailed comparison  between HI and ${\rm He^+}$ line absorption,
and hence a much more
powerful  constraint on models.
We   find significant ${\rm He^+}$ absorption from  HI clouds
(with optical depth of the order of unity)
but also comparable ${\rm He^+}$ absorption
 even in redshift intervals where the best Keck spectrum
reveals no detectable HI; thus we   directly  measure
absorption attributable separately to both the clouds
and the diffuse gas.
Even with high resolution however  our spectrum  suggests
nonzero flux at all wavelengths, which 
constrains the ionizing background spectrum and  leads
to an upper limit on the density of diffuse gas.
Absorption 
from gas near the quasar,  where the incident spectrum is 
known approximately from the direct measurement of the quasar spectrum,
allows independent constraints on the density and helium abundance
of the gas.

\section{Observations and Reductions}

On 3 separate visits in October, November, and December of 1995, Q0302 was
observed with the Goddard High Resolution Spectrograph (GHRS). The G140L
grating and ``D1" detector were chosen especially because of their high UV
efficiency, but this setup also provides good spectral resolution of 0.6\AA\
(about one diode). Q0302 is too faint for a direct target acquisition with
GHRS, necessitating initial acquisitions with the Faint Object Spectrograph,
followed by an offset into the GHRS ``Large Science Aperture" (LSA). This
scheme yields a target-positioning uncertainty in the LSA that potentially
could translate into a $\approx 1$ \AA\ uncertainty in the absolute wavelength 
calibration. (However, empirically we find, for example, an offset of 
only 0.3 \AA\
between the expected and observed wavelengths for the strong interstellar 
C~II absorption component at 1334.5 \AA.)

The requirement to use the LSA results in two complications related to 
limiting geocoronal/airglow contamination. First, a grating tilt was 
chosen with coverage of 1240--1525 \AA\ that starts safely redward of the very
strong geocoronal Lyman-$\alpha$ line. Second, OI 1304~\AA\ airglow cannot 
be avoided completely as its wavelength is very near to redshifted ${\rm He^+}$ 
Lyman-$\alpha$ for Q0302. Hence, the most useful science observations may
be collected only during the spacecraft ``nighttime" portions of each HST 
orbit, and the STScI staff helped insure that scheduling occurred in a manner 
to maximize the availability of such dark time.

Based on the earlier FOC estimates of the UV flux, we expected a net count 
rate from Q0302 significantly lower than the typical GHRS noise background 
rate. Hence, GHRS science data during the nighttime portions of 16 HST orbits 
were taken using special noise-rejection commanding known as 
``FLYLIM", described in greater detail below. In addition, most of 2
additional HST orbits were devoted to routine ``ACCUM" GHRS observations 
of Q0302 during spacecraft nighttime. Finally, there were brief periods 
at the beginnings or ends of each HST target visibility period in which 
the spacecraft was not in Earth shadow; additional GHRS data were collected 
in routine ACCUM mode during these brief spacecraft ``daytime" periods.

Amongst these various datasets, the FLYLIM/nighttime observations have the 
greatest statistical precision due to both their longer total integration time,
plus reduced noise background, and hence are the focus of most of our analysis. 
FLYLIM cut the noise background count rate by nearly one-half while rejecting 
only 7.5\% of the integrations. However, as described below our GHRS 
observations reveal a UV flux for Q0302 in excess of initial FOC 
expectations (and even somewhat higher than a recent re-calibration of the FOC 
data by Jakobsen 1996). This higher UV flux 
interacts with the FLYLIM noise-rejection scheme in a nonlinear
manner that necessitates 
more sophisticated reductions than are standard in the STScI pipeline routines.
The additional 2 orbits of ACCUM/nighttime observations are essential to 
confirming the flux-calibration of the FLYLIM/nighttime observations. The 
ACCUM/daytime observations are strongly contaminated, especially by geocoronal 
emission, and therefore not of much direct use for many of the science issues 
addressed here; nonetheless, they provide useful secondary information on 
airglow line profiles, and on the distribution of background noise events.
\footnote{A preliminary analysis using the OI line profile
collected from the daytime portions of the orbits suggests that OI
contamination is likely to be small in the nighttime observations emphasized
here.}

\subsection{Calibration of FLYLIM Data}

As the proper flux calibration of the FLYLIM/nighttime observations is not
automatically handled in the standard STScI pipeline reductions, we 
discuss here in detail the  relevant offline flux
corrections. The GHRS background noise (mainly cosmic ray induced Cerenkov 
radiation) is non-Poissonian, arriving predominantly in short bursts of events,
and FLYLIM commanding is intended to detect and reject such noise bursts. 
FLYLIM rejects (onboard the spacecraft) any 0.2s
sub-integration in which the total counts accumulated in all 500 GHRS diodes 
equals or exceeds some specified number of counts; in this case a threshold of 
3 counts was selected. This threshold is substantially larger than the count 
rate expected (per 0.2s) from the QSO itself across all 500 diodes,
and thus sub-integrations at or exceeding the threshold of 3 counts are presumed to 
be noise and rejected. Typical GHRS observing modes spend of order 1/16 of 
the actual integration time monitoring the noise background and FLYLIM is 
commanded to reject both on-source sub-integrations (source$+$background) and 
background monitoring (background only) sub-integrations with the identical
threshold.

There is thus the potential with FLYLIM for subtle background and threshold 
rejection effects not accounted for in the standard STScI pipeline flux 
calibration. It is helpful to conceptualize the needed offline flux corrections 
for FLYLIM data
as due to two principal effects, although of course these two are coupled (and 
hence require a coupled calibration/correction).

First, even if there were no background noise, there would be rare (but 
expected) positive counting fluctuations from the source (QSO) that would
occasionally exceed the FLYLIM threshold, hence causing the rejection of an 
entire 0.2s sub-integration. Even though rare, such large rejected positive 
source fluctuations will not be accounted for in the standard pipeline
estimation of the source count rate, and this artificially suppresses the 
standard pipeline inference about the QSO flux. This effect, of course, is 
wavelength (or diode) 
dependent, as the QSO count rate itself is wavelength dependent. This effect 
will be especially important in those regions of the spectrum in which the 
count-rate is highest, as it is in such wavelength/diode regimes where
the largest positive source fluctuations are most likely. (In the case of 
Q0302, this will be longward of the ${\rm He^+}$ break). The flux correction
needed to the pipeline-reduced spectrum for this effect is largely a 
multiplicative one; that is, the actual
flux at each wavelength is (approximately) some 
fixed factor higher than 
that implied by the standard pipeline reductions, and may be simply estimated 
as described below.

A second, and perhaps more subtle effect, is that the background count rate
may also be misinterpreted with FLYLIM. For concreteness, consider a case in 
which source$+$background are being observed, and a background noise event 
generates 2 counts in a particular 0.2s sub-integration while the source 
generates 1 count in that same sub-integration. This 0.2s sub-integration on
source$+$background will be rejected by FLYLIM as it attains the threshold
value of 3. However, when monitoring only the background (no source), there 
would not have been a FLYLIM rejection as the 2 count background event itself 
falls below the rejection threshold. Thus, comparatively lower background rates 
may lead to FLYLIM rejections when observing source$+$background than those 
that lead to FLYLIM rejections when monitoring background only. The result is 
that for observations taken in FLYLIM mode, the noise background count rate 
estimated by the standard pipeline reductions (from background-only monitoring
data) appears {\it higher} than the noise background count rate that is 
actually allowed through when observing both background$+$source.
This latter effect essentially necessitates a zero-point count-rate
(or flux) offset correction
to the pipeline reduced data. This zero-point offset
is especially important for the low count rate portions of the FLYLIM
spectrum (shortward of the ${\rm He^+}$ break for Q0302), and is corrected for as
described below.

\subsection{Empirical FLYLIM Flux Correction}

With the above simple conceptual model in mind, an empirical flux correction 
can be made to the standard pipeline reductions of the FLYLIM/nighttime 
observations, by requiring that the corrected spectrum have both the 
same overall (low-resolution) spectral energy distribution and the same 
mean flux as measured directly from the routine ACCUM/nighttime observations. 
In practice, this is done iteratively. First, an initial estimate of
the mean zero-point offset flux correction (corresponding to the second effect 
discussed in section 2.1) is obtained from the difference between
mean fluxes of the routine ACCUM/nighttime and FLYLIM/nighttime 
pipeline-reduced spectra; the mean fluxes compared here are those shortward of 
the ${\rm He^+}$ break in Q0302, where this zero-point offset dominates. Then, with 
this initial zero-point correction applied, the ratio of mean fluxes between 
the (now partially corrected) FLYLIM/nighttime spectrum and 
the routine ACCUM/nighttime spectrum is determined to estimate the (mainly)
multiplicative correction corresponding to the first effect discussed
in section 2.1; the mean fluxes compared here are those longward of the 
${\rm He^+}$ break, where the first effect dominates. This procedure is iterated 
until the difference in mean fluxes below the break converges to near zero 
(to within $\sim 3\times 10^{-18}$erg/sec/cm$^2$/\AA) and
simultaneously the ratio of mean fluxes longwards of the break converges to 
unity (to within $\sim 0.3\%$). [Note that for these data, in practice it makes 
no significant difference whether such an empirical correction is done in flux 
or count/rate units, although the latter is more rigorously consistent.]

The empirically corrected FLYLIM/nighttime spectrum is displayed in figure 1, 
with the accompanying formal error spectrum. The empirical 
correction in this case may be thought of as roughly equivalent to a zero-point 
offset correction of about 2.5$\times 10^{-17}$ erg/sec/cm$^{2}$/\AA\, coupled 
with a 30\% multiplicative 
correction to the flux. The error spectrum displayed in figure 1 is appropriate 
for assessing the statistical significance of spectral features, but our
empirical correction to bring the FLYLIM/nighttime spectrum into agreement with 
ACCUM/nighttime spectrum also must reflect the zero-point flux uncertainty of 
the  ACCUM/nighttime data. Hence it should also be emphasized that the
empirically corrected FLYLIM/nighttime spectrum of figure 1 has inherited a 
zero-point uncertainty of about 
1.1$\times 10^{-17}$ erg/s/cm$^2$/\AA\ (and this zero-point uncertainty is not
reflected in the formal error spectrum).

On the other hand, it should also be realized that
for most of those discussions below that refer to flux measures--e.g., 
estimates of the optical
depth over broad bands due to ${\rm He^+}$ absorption--one can use {\it either} 
the entirely routine ACCUM/nighttime spectrum or the empirically corrected 
FLYLIM/nighttime spectrum of figure 1. Both spectra yield very similar
results (including errors, which are dominated over broad bands by the 
zero-point uncertainty associated with the 
ACCUM/nighttime observations). 

\subsection{Model FLYLIM Flux Correction}

We have also independently confirmed the FLYLIM/nighttime flux 
corrections by a more detailed quantitative model of the FLYLIM rejection 
process; in contrast to the empirical correction of section 2.2, the model 
correction described now does not depend on any recalibration to the
ACCUM/nighttime flux.
For this more detailed confirmatory model, we assume that the source (QSO)
count rate (per 0.2s sub-integration) may be modeled as a Poisson process. 
For the distribution of GHRS background noise events, a Poissonian model is 
known to be inadequate so we instead estimate this noise count distribution 
(i.e., what fraction of noise bursts yield 0,1,2,3,... count events in each 
0.2s sub-integration) from the background-only monitoring data collected during 
each HST science orbit in routine ACCUM mode (i.e., no source and no 
FLYLIM rejection). These background-only monitoring data 
are essentially background ``spectra" in which all 500 GHRS diodes are 
read-down in exposures lasting about 7 sec each. We rebin each such background 
spectrum into bins of $\approx$14 diodes in width, and count the number 
of background events yielding 0,1,2,3,... counts in each bin. Effectively, 
each bin then samples 14-diodes $\times$ 7 sec $\approx$ 100 diode-secs worth 
of background monitoring data; for the model, we assume that
this binning scheme yields a similar distribution of background events as that 
actually sampled (aboard the spacecraft) in each 0.2s sub-integration across 
all 500 diodes, as it is also the case that 500-diodes 
$\times$ 0.2s $\approx$ 100 diode-secs. Finally, we also require that the model 
account for the empirical constraints that 7.5\% of all 
integrations with FLYLIM implemented were rejected (with the chosen threshold 
of 3), and that 
the pipeline-inferred (i.e., uncorrected) background count rate 
while monitoring background-only (no source) with FLYLIM is 
0.0036 cnts/s/diode. 

The pipeline-reduced FLYLIM/nighttime spectrum is input to the model as
an initial estimate of the actual spectrum, and the model provides a first
estimate of the required flux correction (accounting for both the 
coupled FLYLIM effects described in section 2.1). This partially corrected 
spectrum is then input again into the model to derive an improved flux 
correction, and the procedure iterated until convergence.
It should be emphasized that the ACCUM/nighttime flux does {\it not} in any
fashion enter into this model for the flux correction to FLYLIM data.

This model-corrected FLYLIM/nighttime spectrum agrees very well with the 
empirically-corrected FLYLIM/nighttime spectrum of figure 1.
The difference between the mean fluxes for model- and empirically-corrected
FLYLIM/nighttime spectra is only 1.0$\times 10^{-17}$ erg/s/cm$^2$/\AA, and 
the difference in means shortward of the ${\rm He^+}$ break is even smaller
at 6.0$\times 10^{-18}$ erg/s/cm$^2$/\AA.
In summary, the excellent flux agreement between empirically-corrected and 
model-corrected FLYLIM/nighttime spectra on the one hand, plus the agreement 
between the model-corrected FLYLIM/nighttime spectrum and the entirely
routine (and empirical) ACCUM/nighttime spectrum on the other hand provide 
strong confirmation that the somewhat subtle offline flux corrections 
described in this section are properly accounted for in the spectrum of 
figure 1.

\section{Results and Interpretation}

\subsection{Summary of Absorption Features}

The spectrum clearly confirms the main result of
Jakobsen et al's FOC study: the existence of an absorption
edge of significant optical depth. It also appears that 
the absorption is indeed 
mainly produced by ${\rm He^+}$ 304~\AA\ line absorption
and not by an unrelated low-$z$ HI Lyman limit cloud.
First, this is confirmed by the location of the break,
which in the GHRS data occurs within about 5.3 \AA\ 
shortward of the quasar systemic redshift 
(this may be compared with the $\pm20$\AA\ 
agreement uncertainty window allowed by the 
original FOC data). Scaling the arguments presented
in Jakobsen et al. (1994), the likelihood of a chance
superposition of an HI Lyman limit system (with
column exceeding $10^{18}$cm$^{-2}$) occurring
this close shortward of the QSO systemic redshift is 
only about 0.3\%. Second, in any case our GHRS spectrum
does not reveal  HI Lyman series absorption,
which would have been detectable if a Lyman limit system 
were responsible for the break. 

Most significantly,
we find
significant   coincidence 
in redshift of helium and hydrogen absorption features---
the  ${\rm He^+}$ Lyman-$\alpha$ forest as well 
as the Gunn-Peterson effect.
The portion of the new spectrum of greatest interest
is displayed in figures 2a and 2b,
overlaid with   synthetic spectra, which are   models of
the absorption expected from the clouds   seen in the HI
Lyman-$\alpha$ forest absorption.  Starting with a Keck spectrum of the quasar
(courtesy of A. Songaila, E. Hu and L. L. Cowie), we have fitted  Voigt
profiles (using VPFIT, Webb 1987)
 to all the HI line features down to a threshold column density
of $10^{12} cm^{-2}$ (e.g., a central optical depth of 0.05
for $b=15$km/s),  then used the column densities, Doppler parameters and
redshifts of these clouds to predict the ${\rm He^+}$ absorption, degraded
to the resolution of GHRS.
(The reason for this roundabout modeling, rather than simply 
multiplying the observed HI transmission by a constant factor
[as in e.g. Songaila et al 1995] is to allow us to model either
turbulent or thermal cases, depending for each argument
 on whether maximum or
minimum helium absorption is appropriate).

   Note that several
${\rm He^+}$ absorption  features 
correlate in detail  with features  predicted from  HI; 
the main edge is itself one example of such a feature, caused
by an identifiable complex of HI clouds. 
This edge is 
centered at $z=3.268$, which   is less than
the systemic redshift of  the quasar
but matches that of the clouds.  This is expected since
 the ionizing radiation from   the quasar  reduces
 absorption from the 
 diffuse   foreground gas more than from saturated clouds (Zheng and
Davidsen 1995, Giroux et al 1995)
 The  traditional redshift $z=3.285$ of
Q0302, which would place the edge at 1302\AA\ instead of the observed
1296.4\AA\, is probably quite close to the ``true'' cosmic or 
systemic redshift of the quasar, as nearly the same redshift is 
derived from both Lyman-$\alpha$ and from other, narrower
lines including the  [OIII] 5007 emission line; 
see Espey et al. 1996.
The coincidence (to within $\sim$ 1 resolution element)   of the 
three reddest absorption features
with the HI predictions-- especially the main edge-- suggests that 
the absorption indeed arises from high redshift helium even 
though the edge is not at 
precisely the quasar systemic redshift.

Beyond the main edge, a significant  ``shelf'' of flux
(mean optical depth about 0.8,
level $11.8\pm 1.4\times 10^{-17}$ erg/sec/cm$^2$/\AA) extends
for about another  10\AA, to about 1283\AA\ (20\AA\ 
shortwards of the ``expected'' edge).
We attribute this entire shelf  
to the  fact that in this region  more of the ${\rm He^+}$ is doubly
ionized due to the hard radiation from the quasar
(the ``proximity effect''), greatly reducing
absorption from diffuse gas.

The second edge at 1283\AA\    coincides with the highest HI column
cloud measured, $\log N(HI)= 15.8$, which may even be optically
thick in the ${\rm He^+}$ Lyman limit. 
At shorter wavelengths the flux decreases further beyond the
 second edge.
For broad comparison with the results of other studies, we note that here
the total ${\rm He^+}$ optical depth (which includes both cloud and
diffuse gas contributions) is $\tau_{total}\approx 2.0$, if we compare
the mean flux averaged over the 40\AA\ region shortward of the ``shelf" 
to a similar region longward of the ${\rm He^+}$ break; the 95\% confidence range on 
$\tau_{total}$ is 1.5 to 3.0,
compared with the shelf average of 0.8. Since the contribution
from highly saturated clouds is little affected by the proximity effect,
the change or 
  extra absorption, an optical depth of the order of 
unity, is probably caused by
unsaturated, redshift-space-filling,
 diffuse intergalactic gas. In this sense we confirm
 quantitatively the conclusions of Jakobsen
et al. (1994).

There is some flux detected even near the very bluest portions of the 
spectrum of figure 1. For example, in a 20 \AA\ bin between 
1240 and 1260 \AA\, the mean flux is 5.0$\times 10^{-17}$ erg/s/cm$^2$/\AA\ 
with a formal (photon counting) error of 
$\pm$0.6$\times 10^{-17}$ erg/s/cm$^2$/\AA\ (and the
zero-point offset error of 1.1$\times 10^{-17}$ erg/s/cm$^2$/\AA\ 
discussed in section 2.2). This is at least indicative of
a flux  recovery  towards lower redshift: for example, the mean 
flux in the 1240--1260 \AA\ bin is marginally larger by about 
2.9$\sigma$ than that (2.8$\times 10^{-17}$ erg/s/cm$^2$/\AA\,
with similar errors) in a redder adjacent bin extending over 1260--1280 
\AA.  Note that the zero-point offset error is not relevant when comparing the
difference in flux between these adjacent bins.
Such a flux recovery is expected in many models and
is certainly expected from the results of Davidsen et al.

 The predicted spectra
in figure 2 depend on the adopted column densities and Doppler
parameters of the ${\rm He^+}$. The models displayed in figure 2a simply assume
two constant
ratios for all the clouds,  $\eta\equiv N(He^+)/N(HI) = 20$ 
(corresponding to a spectral slope $\alpha= 1.8$, the
typical value observed in low redshift   quasars and probably 
appropriate for the unabsorbed ionizing
spectrum in the  near-quasar zone here),
and  $\eta = 100$ (corresponding to our
limit $\alpha\ge 3$ derived below for the slope far from the quasar).
Figure 2b shows the result for maximal absorption from an extremely soft
spectrum, in order to demonstrate that the observed HI clouds alone
cannot for any spectrum explain the absorption in the Lyman-$\alpha$
voids.
All the 
 models shown in figure 2 have   $\xi\equiv b_{He}/b_H= 1$ for all components  (pure  
  turbulent broadening, giving the maximal helium/hydrogen optical
depth).
The absorption is quite sensitive to $\xi$ but not to $\eta$ as the helium
absorption by clouds is highly saturated; even very large values
of $\eta$ do not yield significantly more absorption
than the models shown if $\xi=0.5$. Due to variations
in $\xi$ a precise agreement of the simple model with the data
is   not expected even where the spectrum is uniform.

In the following subsections we will develop several
separate quantitative lines of argument based on different features
of the spectrum.
In the ``shelf''  region of the spectrum
the absorption is, within our errors
(including  the uncertainty in the intrinsic quasar spectrum),
accounted for entirely by the gas in clouds,
provided   $b_{He}\approx b_H$   for at least some of the clouds.
Diffuse absorption of   large optical depth in this region is
however not consistent
with the data, and  this fact can be used
together with the known ionizing spectrum of the quasar to
set an upper limit on the diffuse gas density near the quasar.
Knowledge of the
spectrum yields broad constraints on the  absolute value of the helium
abundance.
Shortwards of 1283~\AA,  
 the forest clouds are not sufficient to explain all of the absorption;
indeed there is significant helium opacity
even in
the  well-known ``void'' in the HI forest (Dobrzycki \& Bechtold 1991)
 from which we derive    a lower
limit on  $\eta$ and 
and a constraint on the intergalactic 
ionizing spectrum far from the quasar, which roughly agrees
with the absolute ${\rm He^+}$ ionizing flux estimated from 
the proximity effect.
In spite of the soft spectrum there is   residual
flux throughout the spectrum,
which   gives a separate upper limit on the diffuse gas density
far from the quasar and also indicates that it is likely 
that the intergalactic helium is mostly doubly ionized
by this epoch.

\subsection{Conditions Near the Quasar}

In the region near the quasar, the 
${\rm He^+}$-ionizing flux can be estimated
directly from the observed quasar flux, with several caveats.
If there is dust absorption along the line of sight, the true
continuum level is everywhere 
greater than observed, and the slope of the 
spectrum is shallower;
in addition,  the accumulated Lyman limit absorption
of foreground HI clouds reduces  the observed flux shortwards of 
912 \AA\ in the rest frame (e.g., Vogel and Reimers 1995);
 and the quasar flux may
vary on a timescale long compared with observational 
baseline but short compared to relevant ionization
and recombination rates.
These effects lead to uncalibrated uncertainties
which can only be removed by additional observations; 
variability and absorption   alter the estimate of the 
local ionizing flux, leading to changes in the proximity arguments,
while the unknown intrinsic continuum slope leads to an
uncertainty  the continuum flux and hence in all our
estimated optical depths.  Our approach here is to 
``parametrize our ignorance'' in  correction factors
for these effects. The factors can be constrained using 
other data, for example spectrophotometry of the quasar 
over a larger wavelength range (extending
longwards of  912 \AA\ in the rest frame), or statistical 
surveys of other objects.

We assume  that the flux at the ${\rm He^+}$
ionizing continuum  edge $\nu_i$ ($\lambda_i= $228 \AA\ rest) would if
unabsorbed be
close to the observed flux longwards of 304 \AA.
(The modest equivalent width inferred for the 304 \AA\ emission line
seems to indicate that the
helium-ionizing flux from Q0302 is largely escaping,
and in the 228--304\AA\
region the spectrum is likely to be be fairly flat.)
We allow for the possibility of other sources of absorption
along the line of sight,
such as an accumulation of Lyman continuum absorption,
reducing the flux by a total factor $R^{-1}$; the flux at the continuum edge if 
unabsorbed is then
$f_\lambda\approx 2.6\times 
10^{-16}R\ {\rm erg\ cm^{-2}\ s^{-1} \AA^{-1}}$,
(or $f_\nu \approx 8 \times 10^{-30}R\ {\rm erg\ cm^{-2}\ \ s^{-1} Hz^{-1}}$).
We assume an intrinsic power law spectral energy 
distribution $f_\nu\equiv f_\lambda c
\nu^{-2}
\propto \nu^{-\alpha}.$ 
Zheng and
Davidsen (1995)  estimate that the
intrinsic   $\alpha\approx 1.5$, although $\alpha\approx 2.4 $
has been measured directly and  $\alpha\approx 1.8$
appears to be typical of nearby samples.
For some arguments it is appropriate  to normalize to the
observed flux at HI Lyman-$\alpha$, in which case
it is appropriate to use a different quantity,
the multiplicative difference in the foreground absorption between
the HI and ${\rm He^+}$ Lyman-$\alpha$, 
$R_{304/1216}\approx 4^{2.5-\alpha}$.
For the plausible range of $\alpha$,
$R_{304/1216}$ is likely in the range between 1 and 4.  
Similarly for some arguments it is appropriate to 
use the ratio at the continuum edges, $R_{228/912}$.

At a point along the line of sight at redshift $z_Q-\delta z$, the
quasar spectral density is (Bajtlick et al. 1988)
$$
F_\nu=4f_\nu \delta z^{-2} (1+z)^5[(1+z)^{1/2}-1]^2
$$
if $\Omega=1$, and
$$
F_\nu= f_\nu \delta z^{-2}  z^2 [1+(z/2)]^{1/2}(1+z)^3
$$
if $\Omega=0$, independent of $H_0$; for the present case $z_Q=3.285$,
we find $ (F_\nu/ f_\nu)\delta z^2 = 6.5\times 10^3$ and 6.3$\times 10^3$
respectively.
Expressing the offset   $\delta z$ from the quasar systemic ${\rm He^+}$
Lyman-$\alpha$ redshift
in terms of  observed wavelength offset
 $\delta \lambda= 304\AA \delta z$ from the quasar
${\rm He^+}$ Lyman-$\alpha$, $304\AA\ (1+z_Q) =1302\AA\ $,
  we   estimate the
spectral flux
$$
F_\nu=1.2\times 10^{-23} (\delta\lambda/20\AA)^{-2}
R\ {\rm erg\ cm^{-2}\ \ s^{-1} Hz^{-1}},
$$
the  ionizing photon
flux
$$
F_\gamma\equiv \int_{\nu_i}^\infty d\nu(F_\nu/h\nu)\approx 1.9\times 10^{4}
\alpha^{-1}
(\delta\lambda/20\AA)^{-2} R cm^{-2}sec^{-1}
$$
 and
the ${\rm He^+}$ ionization rate (e.g., Osterbrock 1989)
  $$
\Gamma_{He^+}\equiv\int_{\nu_i}^\infty d\nu(F_\nu\sigma_\nu/h\nu)
\approx 3.0\times 10^{-14} (\alpha+3)^{-1}
(\delta\lambda/20\AA)^{-2} R sec^{-1},
$$
where
the photoelectric cross section of ${\rm He^+}$,
$\sigma_\nu = 1.6\times 10^{-18} (\nu_i/\nu)^{3} cm^2$.

\subsection{Quasar Lifetime and Ionizing Background from the Proximity Effect}

A quasar can doubly
ionize  all the helium  out to a distance of $r=[\delta z/(1+z)] [c/H(z)]$
if the   photon flux $F_\gamma$ is maintained for
a   characteristic time  $t_Q=r n_{He}/3F_\gamma$,
where for cosmic abundance (0.08 by number) the density of helium atoms
is related to the density parameter  of gas
by $n_{He}=9\times 10^{-7}\Omega_g h^2 (1+z)^3 cm^{-3}$.
\footnote{We ignore helium recombinations
for this argument.
In gas with density contrast $\rho/\bar\rho$,
 the recombination rate   is
$
n_e \alpha_{He^+}\approx 8\times 10^{-17}\Omega_g h^2 (\rho/\bar\rho) sec^{-1}
$
which is slower than  the  Hubble rate $H\approx 3\times 10^{-17}h sec^{-1}$
for   $\Omega_g\approx 0.01$. If much of the gas is in density concentrations,
recombinations can be important, requiring a longer quasar lifetime.}
For $\Omega =1$ we
estimate a time for double-ionizing to offset $\delta \lambda$,
$$
t_Q\ge 0.6\times  10^7 h^{-1}\ {\rm yr}\ \alpha R^{-1} (\delta \lambda/ 20\AA)^3
(\Omega_gh^2/10^{-2}),
$$
and about twice this for $\Omega= 0$.
It is therefore plausible for 
 the  helium ionization zone  to extend
about 20\AA\    shortwards of the quasar redshift
 if  the  lifetime of this quasar is
  of the order of
 the characteristic Eddington/Salpeter evolution time for
accretion,  $4\times 10^8\epsilon$yr for typical radiative efficiency
$\epsilon\equiv E_{total}/Mc^2\approx 10\%$,
even if the gas begins
as mostly singly ionized.

If we take $\delta\lambda\approx$20\AA\ as the point where
the quasar ionizing flux is equal to the ionizing background,
  the ionizing spectrum has a specific intensity
$ J_{228}\approx 10^{-24}R\  
{\rm erg\ cm^{-2}\ \ s^{-1} Hz^{-1}sr^{-1}}$, implying
a soft spectrum, with ratio of hydrogen to helium
intergalactic ionizing fluxes $S \equiv J_{912}/J_{228}
\approx 10^3 R^{-1} J_{912,-21}$.

The predictions based on observed quasar populations
are quite sensitive to the assumed typical value of 
$\alpha$ and to models of the absorption. Haardt and 
Madau (1996) predict  $\eta\approx 40$ at this
redshift, although the prediction is likely to increase
as $\alpha$ is revised (from 1.5 say to 1.8;
P. Madau, private communication).  The flux
here corresponds to $\eta\approx 1.7S\approx 200$
for $R=4$ and $J_{912,-21}=0.5$; within the uncertainties
of both  arguments, we consider the agreement satisfactory,
almost remarkable since it is derived from just one quasar.
(Note the independent limit derived below, $\eta\ge 100$,
without reference to the proximity effect.)

\subsection{Diffuse Gas Near the Quasar}

The lack of diffuse absorption leads to an upper limit on the diffuse gas
density
in the $He^{++}$ bubble, since the ionizing  flux from the quasar is known.
Continuous  absorption of optical depth $\tau_{GP}$ is
produced by diffuse ${\rm He^+}$ (which in this
context means atoms producing unsaturated absorption at redshift $z$),
where the density is given by the standard
Gunn-Peterson formula (Peebles 1993)
$$
n_{He^+}= (8\pi/3) \tau_{GP} [\lambda_\alpha (1+z_\alpha)]^{-3} H(z)
\Lambda_\alpha^{-1},
$$
where  the transition rate for helium Lyman-$\alpha$,
$\Lambda_\alpha(He)=16\Lambda_\alpha(H)=1.0\times 10^{10}\  sec^{-1}$.
At $z=3.285$ this becomes $n_{He^+}= 0.9 \times 10^{-9}h \tau_{GP}  cm^{-3}$
or equivalently
 $$
\Omega_gh^2 =1.7\times 10^{-5} h \tau_{GP} (n_{He^+}/ n_{He})^{-1}.
$$
The fraction of helium in ${\rm He^+}$ is given by
$$
{n_{He^+}/ n_{He}}
={n_e\alpha_{He}/ \Gamma_{He^+}}
\approx
3.6\times 10^{-2} {(\rho/\bar\rho)}\Omega_gh^2
(\delta \lambda/ 20\AA)^{2}(\alpha+3)R^{-1} T_4^{-1/2},
$$
    where the density contrast of the material is ${( \rho/\bar\rho)}$ and the
recombination coefficient  is
given near $T_4\equiv T/10^4K=1$ by (Spitzer 1978)
$$
\alpha_{He^+}= 3.4\times 10^{-13}Z^2 T_4^{-1/2} cm^3 sec^{-1}.
$$
This results in a limit on the diffuse  gas density  in the bulk of
velocity space (and hence, the bulk of the intergalactic spatial volume),
$$
\Omega_g h^2=
 2.1\times 10^{-2} \tau_{GP}^{1/2} ( \rho/\bar\rho)^{-1/2} (\delta \lambda/
20\AA)^{-1}
(\alpha+3)^{-1/2}R^{1/2}T_4^{1/4}
h^{1/2}
$$
where $\tau_{GP}$ is the limit on the continuous opacity
at the offset $\delta\lambda$.
We conservatively estimate from our data
$\tau_{GP}\le 1.35$ at $\approx 1285$\AA\
(a 95\% confidence limit, conservatively neglecting any discrete cloud opacity
contribution),
leading to an  upper limit on diffuse intergalactic gas of
$$
\Omega_g\le 0.019 R^{1/2}(\alpha/1.5)^{-1/2}(h/0.7)^{1/2},
$$
tied to the observed quasar flux.  

This limit is interesting since it is of the same order
as the baryon density required by 
Standard Big Bang Nucleosynthesis
(Walker et al 1991, Smith et al 1993, Copi et al 1995, Sarkar 1996,
Hogan 1997).
Current estimates of the total
range from $\Omega_b\approx 0.01(h/0.7)^{-2}$
 (e.g. Rugers \& Hogan 1996) 
 to $\Omega_b\approx 0.05(h/0.7)^{-2}$ (e.g. Tytler et al. 1996).
Realistic CDM models   predict that
most of the baryons should be concentrated in clouds
by this epoch, so our result accords with expectations 
even for large  baryon density (Croft et al. 1997).

\subsection{Helium Abundance}

Since both hydrogen and helium are mostly ionized,
a precise abundance measurement is not possible 
from absorption which only studies a small fraction
of the material. However, because of the unique
information on helium at high redshift
(before the bulk of baryonic material had
even formed into stars), it is interesting   to 
ask how  our data quantitatively constrains
the abundance, even if imprecisely.

An absolute helium abundance can be estimated from
HI and ${\rm He^+}$ Lyman-$\alpha$ absorption 
but only if:
(1)  the helium abundance is uniform; (2)  
 ${\rm He^+}$ and HI are in ionization 
equilibrium; (3)
we know the shape of the ionizing spectrum; 
(4) absorption is unsaturated, so column densities
of absorbing species can measured.
The total columns of   ${\rm He^+}$ and HI are then
both proportional to the same 
line integral $\int d\ell n_e^2$ with coefficients depending
on the abundance and the ionizing spectrum.
(Note that even this statement applies only to
the redshift integrated column densities; there is
only agreement at particular redshift in the case of 
negligible thermal contributions to the atomic velocities.
Note also
that in principle, another test is possible:
in regions where the ionizing spectrum is uniform,
even if it is not known, constraints on 
variations in $\eta$ translate into constraints
on the spatial variations of $Y$.)  

These conditions are certainly not met here in detail,
but the data also do not allow arbitrary 
variations in abundance.
For example, ionization equilibrium
relates the  value of the (redshift-) integrated column density 
ratio $\eta$ to the absolute abundance of helium,
$$
{Y\over 0.24}\approx {\eta\over 1.7S_{228/912}}
$$
provided we know the ratio of ionizing fluxes
$S_{228/912}$.  We can explore the 
range of allowed $\eta$ by our family of models based on
the HI absorption, and the value of $S_{228/912}$
is constrained  in the region dominated by quasar
radiation, subject to the uncertainties discussed above.
Using the above estimates of quasar fluxes, 
$S_{228/912}\approx 28 R_{228/912}^{-1}$,
yielding
$$
{Y\over 0.24}\approx 0.84
\left({\eta\over 10}\right) \left({R_{228/912}\over 4}\right)^{-1}
$$
The best guesses  from the current data
are that $10\le\eta\le20$ and $2\le R_{228/912}\le 4$,
so that $Y$ must lie within   a factor of a few 
of the standard big bang prediction. Although 
the constraints on $\eta$ and $R_{228/912}$
will both improve (the first from better
signal to noise, the second from better
spectrophotometry), it is still unlikely that
a reliable estimate can be made much more precise than
this because of the many assumptions required.
The most interesting new result here is the 
approximate concordance with Big Bang predictions
 at a large distance and an early
epoch, and over a large volume of space.

\subsection{Diffuse Gas in the HI Lyman-$\alpha$ ``Voids'': Lower Limit}

Assuming ${\rm He^+}$ is the dominant species  
leads to a conservative  lower limit on  the density
in absorbing  gas,
$$
\Omega_g h^2 = 1.65\times 10^{-5}h
\tau_{GP}(n(He^+)/n(He))^{-1}[(1+z)/4.285]^{-3/2}.
$$
In the HI Lyman-$\alpha$ void redshift range (especially near
$\lambda_{HI} \approx 4\times$1266\AA), the optical depth 
of ${\rm He^+}$ absorption required by our data,  
after allowing for the absorption from identified clouds
(which in the void is almost independent of $\eta$),  is
still $\tau_{GP}>1.3$ (95\% confidence), requiring  a diffuse
density,
$$
\Omega_g > 3\times 10^{-5}(h/0.7)^{-1}.
$$
This is essentially the same as derived by Jakobsen et al., except that
we can now rule out the possibility of producing this opacity  with HI clouds
down to  the detectability threshold
$\tau_{GP}\le 0.05$ in HI absorption $N(HI)= 10^{12}cm^{-2}$.
The higher resolution here shows that the
helium opacity appears even between 
the most rarefied detected HI clouds. 
Of course for $\eta\ge 100$, much of the ${\rm He^+}$ absorption
could still be from saturated lines not yet resolved in our data.

 In current models,  
absorption is produced in   components
 of lower column density 
but these components
 are produced by ``clouds'' which are indistinguishable from
(are really just parts of) the 
diffuse, space-filling
 protogalactic medium (Cen et al. 1994; Hernquist et al. 1996;
Croft et al 1997; Zhang et al 1997; Bi and Davidsen 1997).
The opacity required by our data is roughly in accord
with these models. We will argue below that it is implausible
to evacuate space with very high efficiency so that    the bulk of the
 helium must be  
doubly
ionized, and even then the constraints on the spectrum impose
an interesting upper limit on the density.

\subsection{Ionizing Spectrum Far from the Quasar}

The large ratio of helium to hydrogen optical depths in the void
  indicates
a   soft ionizing spectrum far from the quasar. If the HI and ${\rm He^+}$ are both
optically thin, the ratio of optical depths 
can be used to constrain $\eta$ directly via
$\tau_{GP}(He^+)/\tau_{GP}(HI)=\eta/4$ 
 (see Miralda-Escud\'e
1993). 
   In ionization equilibrium 
$\eta$ is related to the spectral
softness parameter $S$ (e.g., Giroux
et al. 1995),
$$
\eta={\alpha_{He^+}\over\alpha_{H}}
{\Gamma_{H}\over\Gamma_{He^+}}
{n_{He}\over n_{H}}=1.7 S
$$
where $\alpha$ and $\Gamma$ are the recombination and ionization
rates respectively for the two species, and we have assumed the cosmic
abundance 0.08 and a temperature $T_4=2$. 
 In the Lyman-$\alpha$ void redshift range (i.e., near $\lambda_{HI}
\approx 4\times 1266$\AA), the average optical depth  of diffuse HI absorption
allowed  is at most 0.05. (Since   uniform HI absorption
   could have escaped detection at this level, this  
 limit of 0.05  assumes that there are some
 variations in $\tau(z)$, as expected   from simulations.)
The optical depth of ${\rm He^+}$ is at least 1.3, requiring 
$\eta\ge 100$, $S\ge 63$, and hence an ionizing spectrum of $\alpha>3.0$.
(This limit becomes stronger if the helium absorption is not
from uniform gas or comes from saturated absorbers.)  This is consistent with only soft radiation
having reached this material as expected if 
 ${\rm He^+}$  
absorption is still strongly modifying the emitted spectra of quasars.
The evidence for a soft spectrum is consistent with 
that inferred from SiIV/CIV ratios (Songaila and Cowie 1996,
Savaglio et al 1997), especially considering the sensitivity of
these estimates to details of the spectrum and relative metallicity
(Giroux and Shull 1997). 

We cannot (without a model of the gas
distribution) derive from this data an upper limit to $\eta$
far from the quasar:
 if we assume only the minimal (thermal)
absorption from the clouds, even $\eta\approx 5000$ does not
yield excessive absorption from the clouds.

\subsection{Upper Limit on $z-$filling Gas from the HI voids}

Although we know that there is diffuse gas between the 
forest clouds, we also know there cannot be too much of it 
or else there would be no light getting through, whereas
we have an upper limit of $\tau_{GP}\le 3$ on the mean
optical depth.  Since we also know the ionizing spectrum
is soft,  we can deduce a limit on the
IGM density, tied not to the ionizing flux from the quasar
(as we did above) 
but to the cosmic ionizing flux at the HI Lyman edge, 
$J_{912,-21}$
 (in units of $10^{-21}{\rm erg\ cm^{-2}\ \ s^{-1} sr^{-1} Hz^{-1}}$), 
which has other observational
constraints such as the HI clouds proximity effect 
(eg Madau and Meiksin  1994):
$$
\Omega_g= 0.07 \tau_{GP}^{0.5}  
 \left({1+z\over 4.3}\right)^{-2.25}  h^{-1.5} S^{-0.5}
 J_{912,-21}^{0.5},
$$
which yields  
$$
\Omega_g= 0.018 (\tau_{GP}/3)^{0.5}  
  ( h/0.7)^{-1.5} (\eta/100)^{-0.5}
 (J_{912,-21}/0.5)^{0.5}.
$$  
We thus get    a   conservative limit $\Omega_g=0.018$ by taking
two 95\% limits, one to constrain $\eta\ge 100$ (from the lower limit
$\tau>1.3 $ in the void, above) and one from the upper limit
on the mean total $\tau<3$ everywhere, and
using  
typical estimates $J_{912,-21}\approx 0.5$
(Haardt and Madau 1996, Giallongo et al 1996, but see Cooke
et al 1996).  For a better estimate we should 
allow for the absorption we know is coming from
the HI clouds; from our near-quasar analysis, we guess
that $\tau\approx 1$ from identified clouds, 
allowing only $1.3\le\tau\le 2$ more from
$z$-filling gas.  A reasonable guess for the diffuse gas density
is then $\Omega_g\approx 0.01 ( h/0.7)^{-1.5}$, or even less if
$\eta$ is larger than 100 as suggested by
the proximity effect. The limit is significantly better
than that from studies
from the HI Gunn-Peterson effect, which yield
limits 
$\Omega_g\le 0.2 J_{912,-21}$ at $z=3$
(Giallongo et al. 1992),  and comparable to the 
(more model-dependent) limit 
$\Omega_g\le 0.01$ at $z=4.3$ (Giallongo et al 1994).

\subsection{Ionization History}

Although it is possible that ${\rm He^+}$  is the dominant species in the 
intergalactic gas, the upper limit of 3 on the mean optical depth would allow
in this case at most a density of $\Omega_g\le 7\times 10^{-5}$ 
in diffuse gas--- a number so low that it appears more likely that
the ionization bubble around Q0302  is
a ``proximity effect''---
that is,    double ionized helium is already  predominant
everywhere, and   we are just seeing the region nearest
the quasar with an even higher ionization.
The main features of
our observed spectrum are indeed predicted by models of the proximity
effect
in which a significant contribution to the mean
${\rm He^+}$ opacity comes from the forest clouds (Giroux et al. 1995).
[Note that HS 1700+64
shows no such proximity effect (Davidsen et al. 1996), which is most easily
explained if   the absorption somewhat later at $z=2.72$
is everywhere dominated by saturated lines in clouds.]
The comoving radius today of the observed $He^{++}$ bubble is
$H_0r_0= [(\delta \lambda/20\AA) ]\times 4600$km/sec and
2200 km/sec respectively for open and flat cosmologies. If the Q0302
bubble is typical,
the protogalactic gas has entropy and ionization-state
inhomogeneities of the
order of unity on this
scale.
The scale of the bubble is not negligible compared to the scale over which the 
``recovery'' appears to occur in our spectrum, so it is not clear whether
we are seeing a cosmic trend or merely the history of radiation
percolation (to the redshift of the ``recovery") along this 
line of sight. There is thus strong motivation for obtaining a
high quality spectrum of Q0302 that extends to lower redshifts than
probed by our GHRS data.

\section{Summary}

The absorption observed here is broadly consistent
with the expectations of hierarchical models of structure
formation and with conservative models of cosmic ionization.
It is clear that this type of data will be an important
constraint on models and their parameters, especially 
concerning the most diffuse gas filling the bulk of 
spatial volume.

The main new conclusions from the current data are:
1. The ${\rm He^+}$ Lyman-$\alpha$ forest is detected;
2. The ``diffuse'' (redshift-space-filling)  
medium is also detected, and must have  a low density ($\Omega\le 0.01
(h/0.7)^{-3/2}$) consistent with standard primordial
nucleosynthesis and models of early gas collapse into protogalaxies;
3. The intergalactic ionizing spectrum is soft ($\eta\ge 100$),
although the intergalactic helium is probably mostly doubly ionized
by $z=3.3$; 4. The helium abundance is within a factor of
a few of standard Big Bang predictions, over a large volume
of space at high redshift.

There is clearly a strong motivation to get a spectrum of  other
quasars of comparable quality at the same redshift;
 if we are to draw universal generic conclusions about
cosmic ionization history, it would be prudent   both to
extend our results to lower redshift in Q0302, and to have more
than a single line of sight to check assumptions about
intrinsic quasar properties and uniformity on different sightlines.
In spite of the persuasive checks of the GHRS
calibration, it would also be good to verify that the zero level is 
correct in order to strengthen our limit on the diffuse gas density.
These programs are now underway  with HST/STIS.

\section{Acknowledgments}
We are grateful for advice and encouragement from A. Davidsen, P.
Jakobsen and B. Margon,   for technical support
from  A. Suchkov, A. Berman, C. Leitherer and others at STScI, for the Keck
spectrum from
A. Songaila, E. Hu and L. L. Cowie, and for useful
comments from B. Espey and S. Heap.  This work was
supported at the University of Washington by NASA, and is  based  on
observations with the NASA/ESA Hubble Space Telescope, obtained  at the Space 
Telescope Science Institute, which is operated by AURA, Inc. under NASA 
contract.

\newpage
\figcaption{HST/GHRS spectrum and formal error spectrum
(latter appropriate for assessing the significance of
spectral features), both displayed
at the instrumental resolution of 0.6\AA, using an iterative empirical
flux correction for the nighttime FLYLIM observations.
The correction scheme leaves a residual 1$\sigma$ uncertainty in the
zero level of $1.1\times 10^{-17}{\rm erg\ cm^{-2}\ s^{-1} \AA^{-1}}$.}

\figcaption{A portion of the HST spectrum overlaid with a model spectrum
predicted on the basis of the
model  distribution of HI derived from
a Keck spectrum of the HI Lyman-$\alpha$ forest.
The quasar emission spectrum is fitted   with a 
flat continuum with flux   $2.6\times 10^{-16}
{\rm erg\ cm^{-2}\ s^{-1} \AA^{-1}}$,
plus an ${\rm He^+}$ emission line (centered at 1303\AA, FWHM 4000 km/sec,
and equivalent width of 5\AA).  Ticks indicate the
fitted  HI velocity components from the Keck spectrum. Doppler
parameters and column densities from the fit were
used to predict the  ${\rm He^+}$ absorption spectrum at the GHRS resolution.
Two predictions are shown in figure 2a (upper panel), 
dotted and dot-dash curves corresponding
to  $\eta =20$ and 100 respectively,
both models assuming pure turbulent broadening, $b_{He^+}=b_{HI}$;
figure 2b (lower panel) shows $\eta=500$ (and $\eta=100$ again for comparison).
Note: (1) the  HST and Keck
spectra appear to show corresponding
absorption features near the ${\rm He^+}$ edge;
(2) $\eta=20$ is probably sufficient to explain
the absorption features near the quasar entirely with clouds; (3)
a large
($\tau_{GP}\ge 1.35$) Gunn-Peterson
optical depth is only allowed outside the 
proximity of the quasar, below about 1283\AA;   (4) there is significant
${\rm He^+}$ opacity ($\tau_{GP} > 1.3$) even at the redshift of the conspicuous HI
Lyman-$\alpha$  forest void near 1266\AA;
(5) there is significant nonzero flux even far from the quasar.}
\clearpage
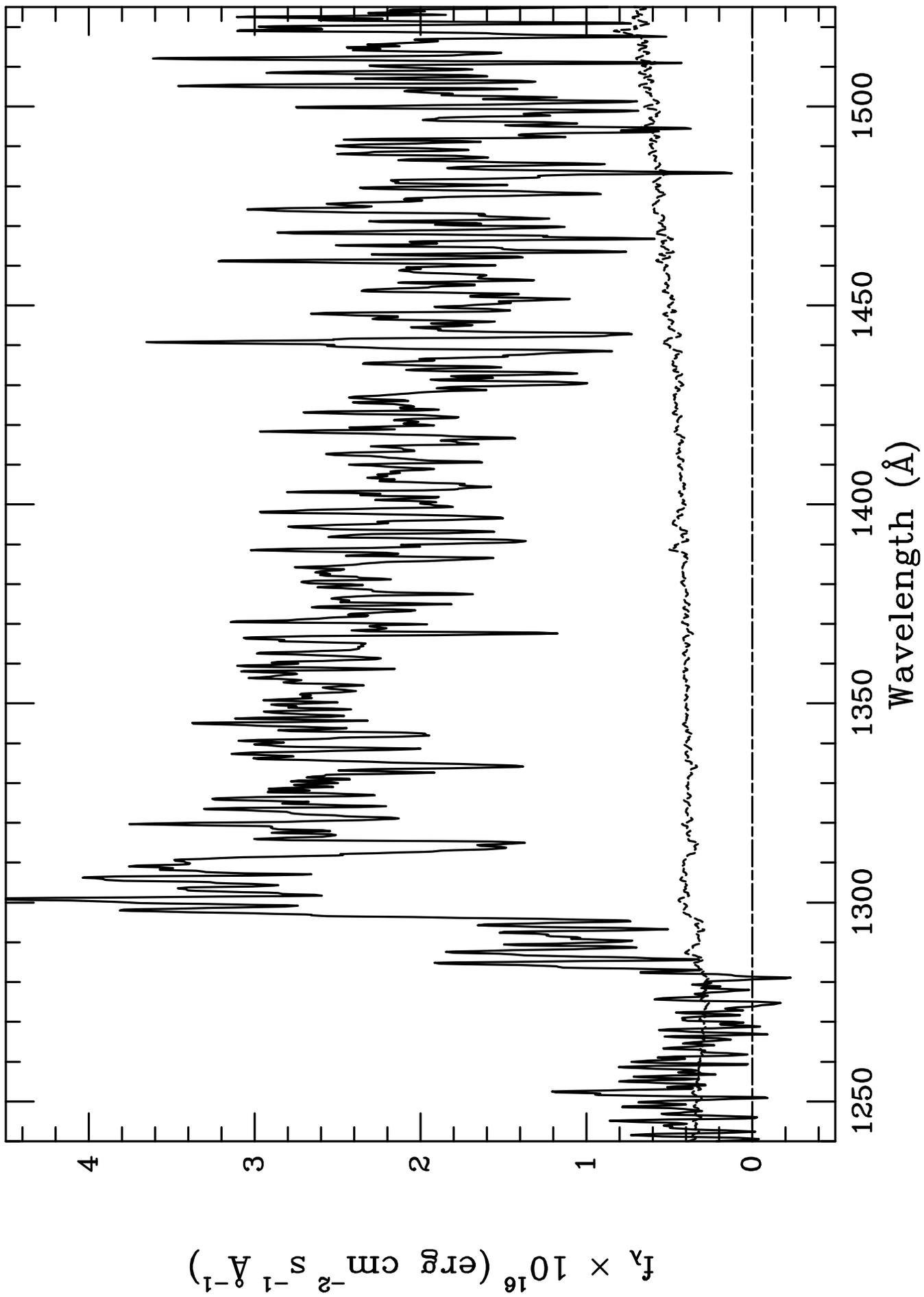
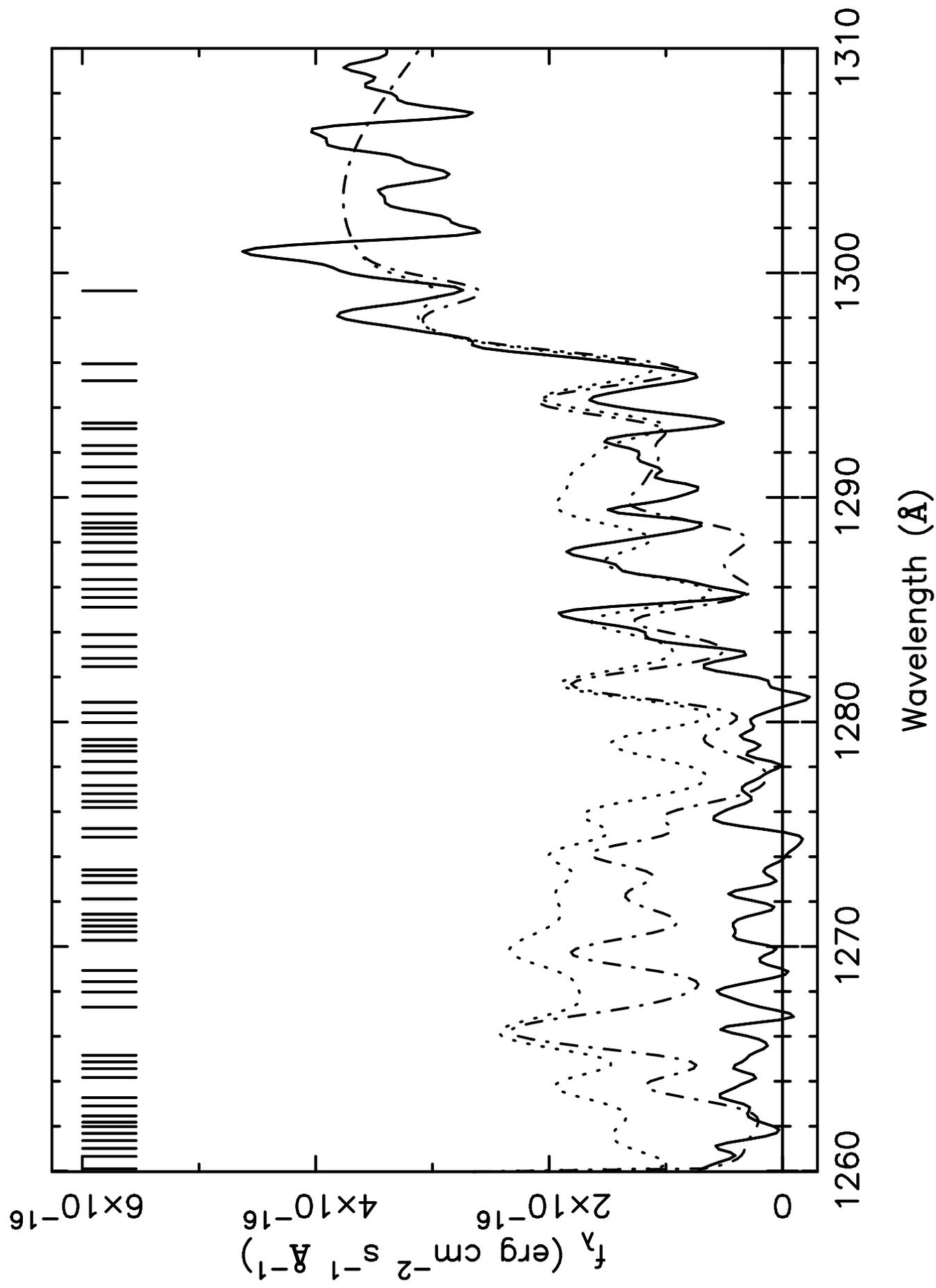
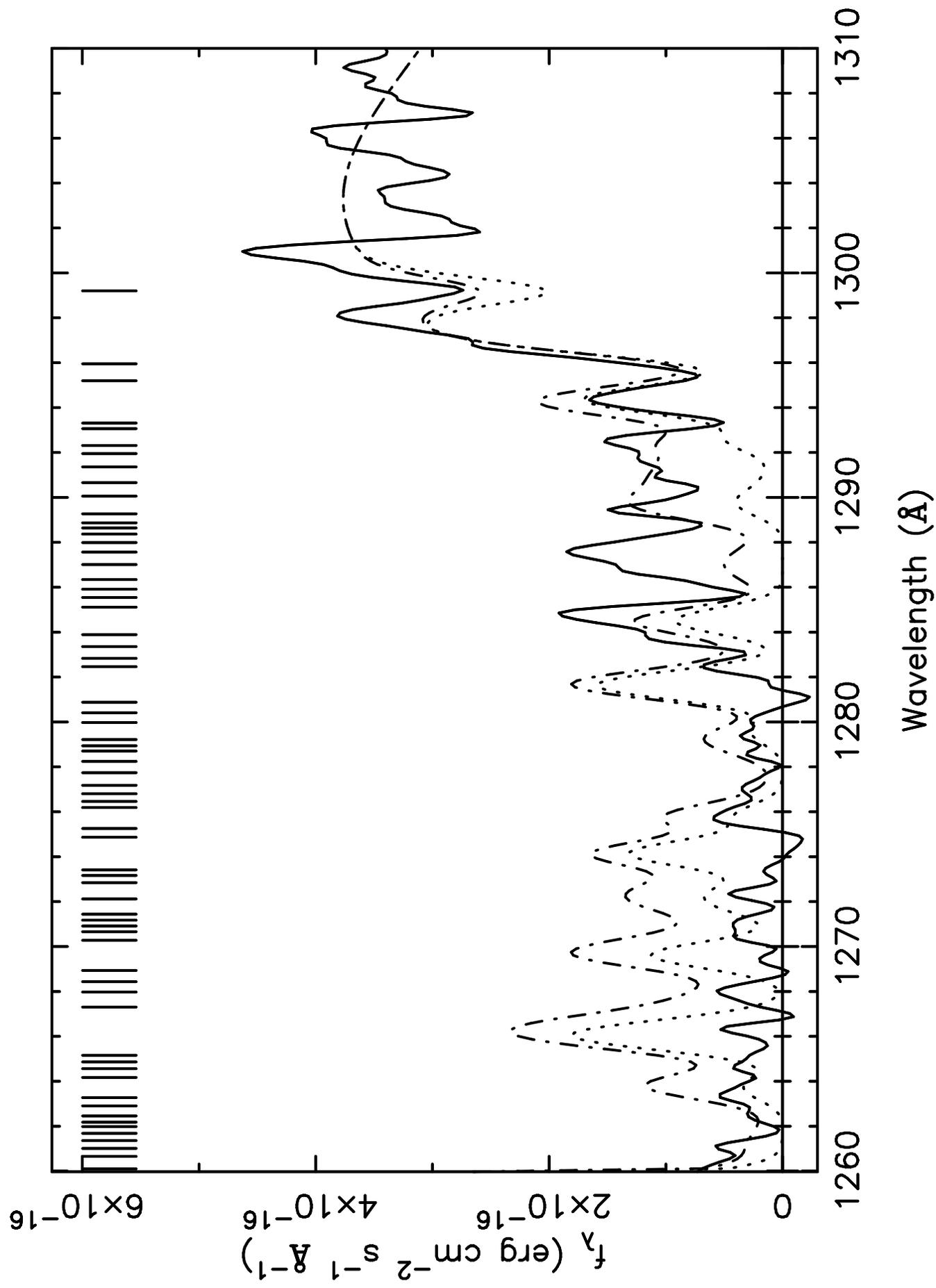
\end{document}